\documentclass{anstrans}
\title{Understanding the Sampling Algorithm for Watt Spectrum}
\author{Jilang Miao,$^{*1}$ and Miaomiao Jin$^{*}$}

\institute{
$^{*}$Department of Nuclear Engineering, The Pennsylvania State University, University Park, 16802 PA, USA
}
\footnotetext[1]{\footnotesize Corresponding author: Jilang Miao (jlmiao@psu.edu)}

\usepackage{graphicx} 
\usepackage{subfigure}

\usepackage{booktabs} 
\usepackage{microtype} 
\usepackage{upgreek}
\usepackage{amsmath}
\usepackage{bbm}
\usepackage{algorithm}
\usepackage{algcompatible}

\DeclareMathOperator*{\argmin}{arg\,min}

\begin{document}
\section{Introduction}
The distribution of fission neutron energy is a crucial step in Monte Carlo simulation for nuclear reactor systems.
The probability density function (PDF) of fission energy  is usually characterized by the Watt spectrum as in Eq.~\ref{eq::watt def}~\cite{brown2016monte}.
\begin{equation}
  f(x) = \frac{2e^{-ab/4}}{\sqrt{\pi a^3 b}} e ^{-x/a} sinh\sqrt{b x}
   \equiv A e ^{-x/a} sinh\sqrt{b x}
\label{eq::watt def}
\end{equation}

The algorithm for sampling the Watt spectrum is widely recognized and is provided in Algorithm~\ref{alg::watt} for completeness. This algorithm has been incorporated into Monte Carlo software packages such as OpenMC~\cite{romano2013openmc} and MCNP~\cite{brown2002mcnp}. However, the rationale behind the selection of parameters $K$, $L$, and $M$ in Algorithm~\ref{alg::watt} remains unclear, and previous research has not provided a detailed explanation for these choices.
\begin{algorithm}
  \caption{Rejection method to sample Watt spectrum.}
  \label{alg::watt}
\begin{algorithmic}
\State Define $K=1+ab/8$, $L=a\left( K + \sqrt{K^2-1}\right)$, $M=L/a-1$
\State Sample random variables $\xi_1$, $\xi_2$ uniformly distributed within $(0,1)$
\State $x \gets -log \xi_1$, $y \gets -log \xi_2$
\IF {$\left( y-M(x+1) \right)^2 \le bLx$}
\State accept, return $Lx$
\ELSE
\State reject
\ENDIF
\end{algorithmic}
\end{algorithm}

The Algorithm ``R12" from ``3rd Monte Carlo Sampler" by Everett \& Cashwell~\cite{everett1972monte} is commonly cited as the source for Algorithm~\ref{alg::watt}. However, it's important to note that Everett \& Cashwell's work provides the algorithm itself without offering detailed derivations or explanations. Additionally, Algorithm ``R12" in Everett \& Cashwell's work references "Computing Methods in Reactor Physics" by Kalos~\cite{greenspan1972computing} for further information, but unfortunately, Kalos' work also presents a version of Algorithm~\ref{alg::watt} without providing the necessary details.
In particular, Kalos' work utilizes a different convention and focuses solely on the specific values of $a$ and $b$ ($a=0.965$ and $b=2.29$) without elaborating on the method of how the values of $K$, $L$, and $M$ are determined. Nevertheless, it is mentioned in Kalos' work that these numerical values were chosen to optimize the efficiency of the rejection method, resulting in an efficiency value of 0.74. Therefore, in this summary, we aim to bridge this gap by deriving the Watt spectrum sampling algorithm and providing a theoretical rationale for the optimal choices made within it.

\section{Methodology}
\subsection{Background}
We first review the rejection sampling method to be used.
Suppose we have a target $1D$ PDF $f(x)$, a $2D$ PDF $g(x,y)$ and $1D$ functions $h_1(x),h_2(x)$, which are convenient to evaluate.
We then assume, the functions $f(x), g(x,y)$, $h_1(x)$ and $h_2(x)$ satisfy Eq.~\ref{eq::fghL}.
\begin{equation}
f(x) = \Lambda \int_{h_1(x)}^{h_2(x)} g(x,y) dy
\label{eq::fghL}
\end{equation}
where $\Lambda$ is a constant needed to normalize $f(x)$ to be a PDF.
Suppose we have two random variables $(\xi_1,\xi_2)$ jointly sampled from PDF $g(x,y)$,
it can be shown that if we accept $\xi_1$ when $h_1(\xi_1) \le \xi_2  \le h_2(\xi_1)$,
$\xi_1$ follows the distribution of $f(x)$~\cite{robert1999monte}. 
The general rejection algorithm is given in Alg.~\ref{alg::reject},
and the efficiency of the rejection method is
\begin{equation}
  \eta = \frac{1}{\Lambda}
\label{eq::eff}
\end{equation}

\begin{algorithm}
\caption{A Rejection method to sample $1D$ PDF $f(x)$.}
\label{alg::reject}
\begin{algorithmic}
\State select function $h_1(x)$,$h_2(x)$ and $2D$ PDF $g(x,y)$ such that are related with target PDF $f(x)$ as in Eq.~\ref{eq::fghL}
\State Sample random variables $(\xi_1$, $\xi_2)$ from PDF $g(x,y)$
\IF {$h_1(\xi_1) \le \xi_2  \le h_2(\xi_1)$}
\State accept, return $\xi_1$
\ELSE
\State reject
\ENDIF
\end{algorithmic}
\end{algorithm}

\subsection{Application}
We then select the corresponding functions $h_1(x)$, $h_2(x)$ and $g(x,y)$ for the target PDF $f(x)$, i.e., the Watt spectrum in Eq.~\ref{eq::watt def}.
By definition of $sinh(x)$, we first rewrite $f(x)$ as
\begin{equation}
\begin{aligned}
  f(x) &= \frac{A}{2}e ^{-x/a} \left(  e^{\sqrt{b x}} -  e^{-\sqrt{b x}} \right)  \\
  & = \frac{A}{2} e ^{-x/a + z(x)}  \int_{z(x)-\sqrt{b x}}^{z(x)+\sqrt{b x}} e^{-y} dy
\label{eq::f+z}
\end{aligned}
\end{equation}
Eq.~\ref{eq::f+z} holds for any function $z(x)$. For simplicity, we can choose a linear form for $z(x)$,
\begin{equation}
z(x) = c_1 x + c_2
  \label{eq::zx}
\end{equation}
where $c_1$ and $c_2$ are constants. With this $z(x)$, we can plug it in Eq.~\ref{eq::watt def},
\begin{equation}
  f(x)   = \frac{A}{2} e ^{-x(1/a-c_1) +c_2 }  \int_{z(x)-\sqrt{b x}}^{z(x)+\sqrt{b x}} e^{-y} dy
\label{eq::f+zc}
\end{equation}

Then two exponential distributions in Eq.~\ref{eq::f+zc} can be identified, one in the leading $e^{-x/a+z(x)}$ term, and
the other in the integrand $e^{-y}$.
Hence, these two PDFs can be merged to create the joint distribution $g(x,y)$, and thus, 
\begin{equation}
  f(x)   = \Lambda  \int_{z(x)-\sqrt{b x}}^{z(x)+\sqrt{b x}} \left(\frac{1}{a} - c_1 \right) e^{-\left(\frac{1}{a}-c_1\right) x}e^{-y} dy
\label{eq::f+g}
\end{equation}
From Eq.~\ref{eq::f+g}, we can identify all the functions required to execute Alg.~\ref{alg::reject}.
\begin{align}
  g(x,y) & = \left(\frac{1}{a} - c_1 \right) e^{-\left(\frac{1}{a}-c_1\right) x}e^{-y} \label{eq::def g}\\
  h_1(x) & = z(x)-\sqrt{bx} \\
  h_2(x) & = z(x)+\sqrt{bx} \\
  \eta &= \frac{1}{\Lambda} = \frac{2}{A} \frac{\frac{1}{a} - c_1}{e^{c_2} } \label{eq::etac}
\end{align}
Note that the $2D$ PDF $g(x,y)$ (Eq.~\ref{eq::def g}) is simply the product of two PDFs. Hence, we can independently sample $\xi_1,\xi_2$ from exponential distribution with parameter $1$ and $\frac{1}{a} - c_1$, respectively.

\subsection{Efficiency Optimization}
We then find the coefficients $c_1,c_2$ in the definition of $z(x)$ (Eq.~\ref{eq::zx})
to maximize the efficiency $\eta$ defined in Eq.~\ref{eq::etac}.
The optimization problem with constraints can be formulated as Eq.~\ref{eq::opt}.
\begin{equation}
  \begin{aligned} 
   c_1^*,c_2^*  = 
    &\argmin_{\substack{\{c_1,c_2\} \\ \frac{1}{a}-c_1>0 \\ c_1 x + c_2 -\sqrt{bx} \ge 0 ~,~ \forall x> 0 }}  \frac{A}{2} \frac{e^{c_2} }{\frac{1}{a} - c_1}
      \label{eq::opt}
\end{aligned}
\end{equation}
To make sure the $x$ distribution as specified in $g(x,y)$ (Eq.~\ref{eq::def g}) is distributed in the positive range, the following constraint needs to be enforced.
\begin{equation}
  \frac{1}{a}-c_1>0
\label{eq::con1}
\end{equation}
Additionally, another constraint as specified in Eq.~\ref{eq::con2} is necessary, which enforces the integral limits $z(x)\pm\sqrt{bx}$ fall within the range of random variable $y$ in Eq.~\ref{eq::fghL} and Eq.~\ref{eq::f+g}.
Since $\sqrt{bx}$ is positive, the constraint on the upper bound is not necessary.
\begin{equation}
c_1 x + c_2 -\sqrt{bx} \ge 0 ~,~ \forall x> 0
  \label{eq::con2}
\end{equation}
Note that, we can shift the range of exponential distribution $e^{-y}$ in $g(x,y)$ from $(0,\infty)$ to $(-S,\infty)$, with $S$ being a positive number, to relax the constraint by Eq.~\ref{eq::con2}.
However, it will not increase the sampling efficiency, as the shift factor $S$ will enter all equations with $c_2$ and
yield results equivalent to that from Eq.~\ref{eq::con2}.

Next, we rewrite the constraint in Eq.~\ref{eq::con2} so that it depends on $c_1$ and $c_2$ only.
$c_1 x + c_2 -\sqrt{bx}$ is quadratic function of $\sqrt{x}$.
Hence, Eq.~\ref{eq::con2} can be equivalently converted to
\begin{align}
c_1 & >0
  \label{eq::con21} \\
 c_2 & \ge \frac{b}{4 c_1}
  \label{eq::con22}
\end{align}

\begin{figure}[htbp]
\centering
\includegraphics[width=0.43\textwidth]{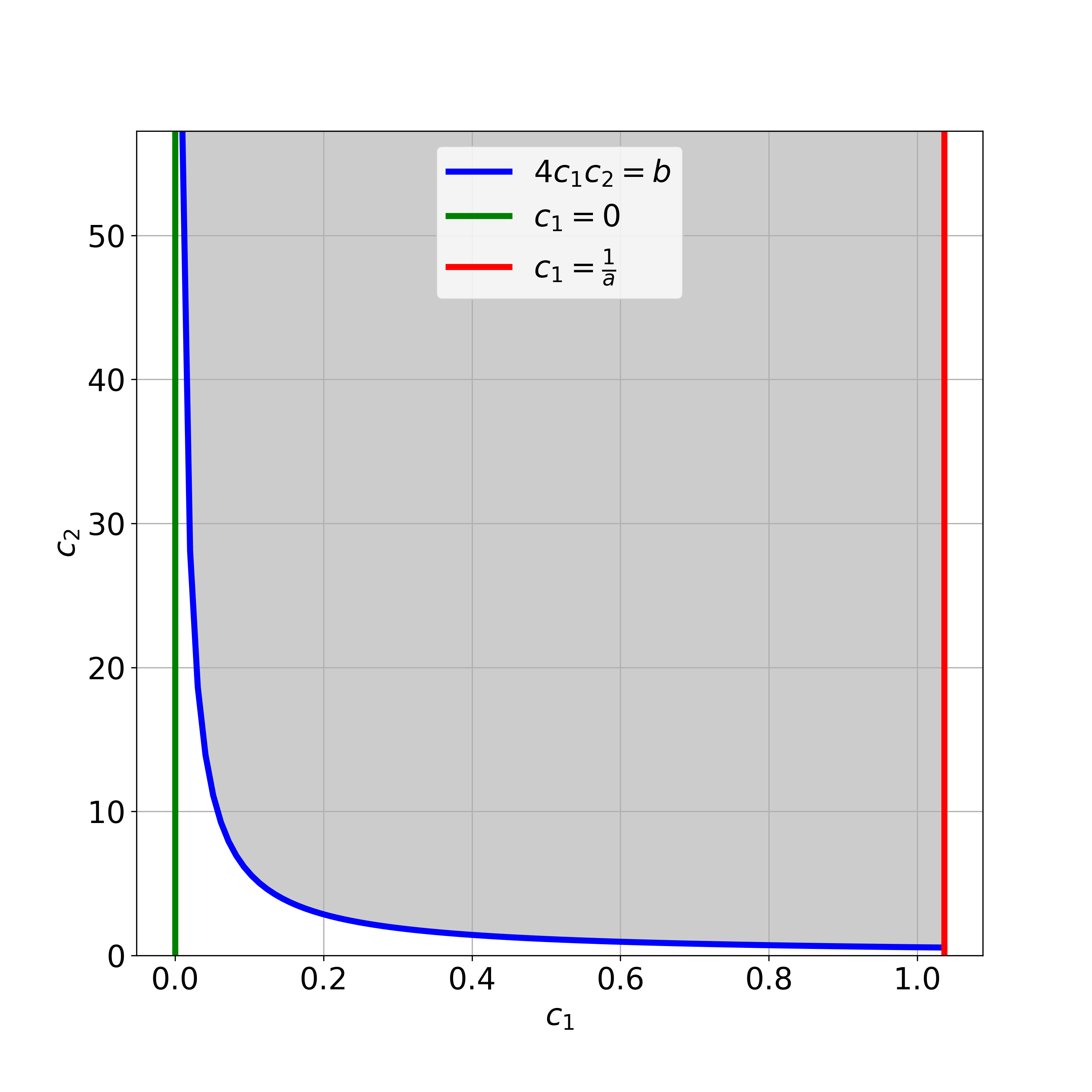}
\caption{Visualizing optimization problem in Eq.~\ref{eq::opt2}. }
\label{fig::opt}
\end{figure}

Now we can simplify the optimization problem from Eq.~\ref{eq::opt} to Eq.~\ref{eq::opt2}.
\begin{equation}
  \begin{aligned} 
   c_1^*,c_2^*  = 
    &\argmin_{\substack{\{c_1,c_2\} \\ c_1> 0 \\  \frac{1}{a}-c_1 > 0 \\ 4 c_1  c_2 \ge b}}   \frac{e^{c_2} }{\frac{1}{a} - c_1}
    =
    \argmin_{\substack{\{c_1,c_2\} \\ c_1> 0 \\  \frac{1}{a}-c_1 > 0 \\ 4 c_1  c_2 \ge b}}  \mathcal{L}(c_1,c_2)
      \label{eq::opt2}
\end{aligned}
\end{equation}
When defining the Lagrangian $\mathcal{L}(c_1,c_2)$, Eq.~\ref{eq::opt2} also discarded the leading term $\frac{A}{2}$ since it does not depend on $c_1,c_2$. 
The optimization problem can be visualized in Fig.~\ref{fig::opt},
where Lagrangian $\mathcal{L}(c_1,c_2)$ in the shaded area would be minimized.

Since we have
\begin{align}
  \frac{\partial \mathcal{L}(c_1,c_2)}{\partial c_1} & > 0 \\
  \frac{\partial \mathcal{L}(c_1,c_2)}{\partial c_2} & > 0 
\end{align}
the optimal efficiency prefers small $c_1$ and $c_2$,
and thus the optimal $(c_1^*,c_2^*)$ must be located on the curve $4c_1c_2=b$. 
This allows us to further simplify the optimization problem with inequality constraints to one with equality constraints. 

\begin{equation}
   c_1^*,c_2^*  = 
    \argmin_{\substack{\{c_1,c_2\} \\ 4 c_1  c_2 = b}}  \mathcal{L}(c_1,c_2)
\label{eq::opt3}
\end{equation}
Eq.~\ref{eq::opt3} can be solved by defining another Lagrangian with multiplier $\mu$ as in Eq.~\ref{eq::lag mu},
and then setting the derivative with respect to the parameters equal to 0. 
\begin{equation}
\mathcal{L}(c_1,c_2,\mu) \equiv \mathcal{L}(c_1,c_2) - \mu (4c_1c_2 - b )
\label{eq::lag mu}
\end{equation}
Therefore, the optimal coefficients $(c_1^*,c_2^*)$ is the solution of the  system of equations in Eqs.~\ref{eq::sys1}~\ref{eq::sys2}~\ref{eq::sys3}.

\begin{align}
  0 &= \frac{\partial \mathcal{L}(c_1,c_2,\mu)}{\partial c_1}  = \frac{e^{c_2} }{ \left( \frac{1}{a} - c_1 \right) ^ 2 } -4\mu c_2
\label{eq::sys1}
  \\
  0 &= \frac{\partial \mathcal{L}(c_1,c_2,\mu)}{\partial c_2}  = \frac{e^{c_2} }{  \frac{1}{a} - c_1  } -4\mu c_1
\label{eq::sys2}
  \\
  0 &= \frac{\partial \mathcal{L}(c_1,c_2,\mu)}{\partial \mu}  = b-4c_1c_2
\label{eq::sys3}
\end{align}

Finally, we find the optimal coefficients as
\begin{align}
  c_1^* &= \frac{2b}{aJ} \label{eq::c1} \\
  c_2^* &= \frac{aJ}{8} \label{eq::c2}
\end{align}
where
\begin{equation}
  J = \sqrt{b \left(\frac{16}{a} + b \right) } + b 
  \label{eq::def J}
\end{equation}
With the optimal coefficients $(c_1^*,c_2^*)$ in Eq.~\ref{eq::c1} and Eq.~\ref{eq::c2}, Alg.~\ref{alg::reject} is found to be identical to the ``R12'' algorithm (Alg.~\ref{alg::watt}) by recognizing the following relationships,
\begin{align}
  c_1^* &= \frac{M}{L} \\
  c_2^* &= M \\
  h_1(x) \le y \le h_2(x) & \Leftrightarrow \left( c_1^*x+c_2^* -y \right) ^2 \le bx
\end{align}

It can also be confirmed that the efficiency of the algorithm (Eq.~\ref{eq::etac}) is equal to $0.7595$ for the case where $a=0.965$ and $b=2.29$.
This value is slightly larger than the value of $0.74$ reported in Kalos' work~\cite{greenspan1972computing} but can be accurately verified from sampling experiments.

\subsection{Alternative $z(x)$ options}
In addition to the choice of a linear function of $z(x)$ as in Eq.~\ref{eq::zx}, $z(x)$ in the following form was also explored. 
\begin{equation}
z(x) = c_1 \sqrt{x} + c_2
  \label{eq::zx2}
\end{equation}
Following a procedure slightly different from above, we will find the optimal coefficients $(c_1^*,c_2^*)$ as 
\begin{align}
  c_1^* &= \frac{\sqrt{b}}{2}(1-J_2) \label{eq::c1 2} \\
  c_2^* &= \frac{(1+J_2)^2}{J_1} \label{eq::c2 2}
\end{align}
where
\begin{align}
  J_1 &= \frac{16}{ab} 
  \label{eq::defJ1} \\
  J_2 &= \sqrt{1+J_1}  \label{eq::defJ2}
\end{align}
However, the optimal efficiency which can be described in Eq.~\ref{eq::eff 2}, is $0.5588$ for the case where $a=0.965$ and $b=2.29$.
\begin{equation}
    \eta = \frac{1}{A} \frac{{c_1^*}^2}{e^{c_2^*}}
    \label{eq::eff 2}
\end{equation}
Eq.~\ref{eq::eff 2} was also numerically verified with Monte Carlo simulation experiments. 

\section{Conclusions }
In this summary, we have re-derived the ``R12" algorithm for the Watt spectrum, with a focus on optimizing the efficiency of the rejection sampling method. This bridges the gap in the algorithm, and its interpretation serves as a motivation for enhancing the Watt spectrum sampling technique.
\section{Acknowledgments}
This work is supported by the Department of Nuclear Engineering, The Pennsylvania State University.

\bibliographystyle{ans}
\bibliography{bibliography}
\end{document}